\documentclass{aastex63}

\newcommand{\be}{\begin{equation}}
\newcommand{\ee}{\end{equation}}
\newcommand{\appropto}{\mathrel{\vcenter{
  \offinterlineskip\halign{\hfil$##$\cr
    \propto\cr\noalign{\kern2pt}\sim\cr\noalign{\kern-2pt}}}}}
\received{}
\revised{}
\accepted{}
\submitjournal{AJ}

\shorttitle{Mutual detectability: avoiding the SETI Paradox}
\shortauthors{Kerins}
\graphicspath{{./}{figures/}}

\begin{document}

\title{Mutual detectability: a targeted SETI strategy that avoids the SETI Paradox}

\correspondingauthor{Eamonn Kerins}
\email{Eamonn.Kerins@manchester.ac.uk}

\author[0000-0002-1743-4468]{Eamonn Kerins}
\affiliation{Jodrell Bank Centre for Astrophysics,\\ 
Dept of Physics and Astronomy, \\
University of Manchester, Oxford Road, \\
Manchester M13 9PL, UK}



\begin{abstract}
As our ability to undertake Searches for Extraterrestrial Intelligence 
(SETI) grows, so does interest in the controversial endeavour of 
Messaging Extraterrestrial Intelligence (METI). METI proponents point to 
the SETI Paradox -- if all civilisations refrain from METI then SETI is 
futile. I introduce {\em mutual detectability} as a game-theoretic 
strategy to increase the success potential of targeted SETI. Mutual 
detectability comprises four laws that establish how SETI participants 
can engage each other based on mutual evidence of mutual existence. I argue that the party whom 
both SETI participants can judge to have better quality evidence, or 
common denominator information (CDI), has an ``onus to transmit'' to avoid the SETI Paradox. 
Transiting exoplanets 
within the Earth Transit Zone form a target subset that satisfies mutual 
detectability requirements. I identify the 
intrinsic time-integrated transit signal strength, which for Earth is 
$10^3$~L$_{\odot}$~ppm~hours~yr$^{-1}$, as suitable CDI. Civilisations on habitable-zone planets of radius $R_{\rm p}/R_{\oplus} 
\lesssim (L_*/L_{\odot})^{-1/7}$ have superior CDI 
on us, and so under the mutual detectability framework have game-theory incentive (onus) to transmit. Whilst 
the onus to transmit falls on us for habitable planets 
around $L_* > L_{\odot}$ stars, considerations of relative stellar 
frequency, main-sequence lifetime and planet occurrence rates mean that 
such systems are likely to be in a small minority. Surveys 
of the Earth Transit Zone for Earth-analogue transiting planets around sub-solar 
luminosity hosts would facilitate targeted-SETI programs for civilisations who have game-theory incentive to transmit signals to us. A choice to remain silent, by not engaging in METI 
towards such systems, does not in this case fuel concerns of a SETI 
Paradox.
\end{abstract}



\section{Introduction} \label{sec:intro}

 A new impetus in the search for extraterrestrial intelligence (SETI) has been provided by increases in detector sensitivity, advances in data science and philanthropic financial backing. These, together with advances in exoplanetary detection and characterisation, bring greater hope that we can answer the question of how common  intelligent life may be in our Galaxy \citep{Wright2017}. Recent results from the Breakthrough Listen project, and other SETI surveys, are placing much stronger limits on extraterrestrial radio transmissions and other forms of technosignature than was possible only a short time ago \citep[e.g.][]{2013ApJ...767...94S,2019AJ....157..122P,2020AJ....160...29S,2020AJ....159...86P,2020MNRAS.tmp.2570W,2020ApJ...891..174Z}.

SETI research has traditionally been regarded as outside of mainstream science activity, as an extreme example of a ``high risk -- high gain'' endeavour where the probability of ``failure'' is seen as almost inevitable. This view is now changing, partly because of the vast increase in capability of current surveys, but also because of an appreciation of the value of such surveys to time-domain astrophysics. SETI surveys can provide the ability to systematically cover regions of the time-domain discovery space that are not easily accessible to other surveys \citep[c.f.][]{2020arXiv200611304L}. Rather than ``failure'' being almost inevitable, SETI surveys have a very high likelihood of success if we choose to use the term ``success'' as it is normally used in observational astrophysics, to refer to the discovery of new phenomena or to expand our understanding of existing phenomena through new ways of observing.

Current survey strategies include both wide-area SETI surveys and also targeted-SETI surveys that are focused on specific areas of sky, or even specific stellar systems. As the potential search parameter space is huge these different strategies have complementary strengths. What  targeted-SETI programs lack in areal coverage can be made up for by sampling a greater range of frequencies. Alternatively, one can undertake more sophisticated signal processing searches that may not be feasible on the much larger data volumes generated by wide-area surveys covering the same frequencies.

The potential for exoplanets located within the Earth's Transit Zone (ETZ) for targeted SETI surveys has long been recognised \citep{10.1007/3-540-54752-5_225,2004IAUS..213..409S,2016AsBio..16..259H,Dogaru2019}. Civilizations on planets within the ETZ are able to observe the Earth in transit against the Sun and therefore discover that Earth is a rocky planet located within the Sun's habitable zone (HZ). With observing capabilities a little ahead of our own they may detect that Earth's atmosphere exhibits biosignatures, or even technosignatures in the form of industrial atmospheric pollutants \citep{2014ApJ...792L...7L}. Any of these types of evidence might well motivate them to listen out for communication signals from us or even to direct signals at us.

As SETI survey activities expand, so does interest in the more controversial topic of messaging extraterrestrial intelligence (METI). Arguments have been put forward for and against the idea of deliberately sending communication signals out into the cosmos in the hope of them being intercepted \citep[for examples of two recent discussions on either side of the debate see][]{2016JBIS...69...31G,2020arXiv200601167C}. One of the principal arguments for engaging in METI has been dubbed the SETI Paradox \citep{2006physics..11283Z}. The SETI Paradox contends that if all SETI-capable civilisations decide that METI is unwise then no transmissions will be made, leading to the inevitable failure of SETI. The idea that SETI cannot succeed without METI places an onus on all civilizations wishing to engage in SETI to also engage in METI.

In this paper I examine the how game theory may be used both to enhance the success potential for targeted SETI and also to consider the question of whether the incentive for METI is shared equally, regardless of questions concerning potential dangers. The implicit role of game theory in SETI has recently been made explicit by \cite{Wright2017,2020arXiv200801817W}. He points out that SETI is a good example of a class of games that are coordination games between two cooperative but non-communicating participants. In such games considerations of incentives, constraints and predicted behaviour motivate the adoption of favoured choices by the participants, referred to as Schelling points. Schelling points can therefore provide a higher than random chance of a successful outcome. 

This paper is structured as follows. In Section~\ref{sec:mutual} I propose {\em mutual detectability} as a principle by which SETI-capable civilisations may engage in targeted SETI in order to enhance the chance of successful communication. The principle comprises four laws, the last of which enables a decision on who has the greater game-theory incentive to engage in METI. The principle of mutual detectability highlights that, since game-theory incentive may not be equally shared, the prospect of the ``Great Silence'' raised by the SETI Paradox represents an unstable solution and should be unlikely to occur if SETI-capable civilizations are common. Section~\ref{transit} considers the specific example of targeted SETI towards transiting HZ planets located in the ETZ to illustrate how the principle can be applied. I show that civilizations on HZ planets located around sub-solar luminosity hosts have game-theory incentive to transmit to us. From a SETI perspective, a good strategy for us would therefore be to undertake a transit survey for such planets and then engage in targeted SETI towards them. For this specific subset of systems at least, we do not have a game-theory incentive to engage in METI. The paper is summarised in Section~\ref{discuss}.

\section{Mutual detectability} \label{sec:mutual}

Mutual detectability is a game-theoretic approach aimed at enhancing the chance of communication between two SETI-capable civilisations. 
I shall assume in this discussion that life with the potential for communication is sufficiently common that detection strategies are amenable to probabilistic consideration. If they aren't then SETI itself is probably futile.

The basic idea underpinning mutual detectability is that the more SETI-capable civilisations know about each other's environment, the more likely they are to attempt communication. Mutual detectability considers the situation where both parties are able to recognize that both of them can access affirmative information on the potential habitability of the other's environment. This situation provides a Schelling point for targeted SETI and motivates the first and second laws of mutual detectability:

\begin{itemize}
    \item {\em 1. Mutuality} -- The potential for verifiable communication between two SETI-capable civilizations is improved wherever {\em mutual} evidence points to their mutual existence.
\end{itemize}

\begin{itemize}
    \item {\em 2. Symmetry} -- The presence of {\em symmetrical} evidence, where the same type of information is mutually available, enables both parties to evaluate the relative strength of the other's evidence. 
\end{itemize}

There are two desirable objectives when considering mutual detectability. One is to have access to information about a target location that can affirm its potential for life, since this provides greater incentive to engage in SETI towards it. The second is that the available information should be as accessible as possible in order to encourage interest from civilizations with a broad spectrum of technological capability. There may be a degree of tension between these objectives in a case where one party is significantly more advanced than another. The more advanced civilization may be able to access a wide array of information about the other civilization that encourages the advanced civilization to attempt contact. However, the less advanced civilization may not appreciate the weight of evidence of their own existence to the other civilisation, and therefore not may not engage. Other targets may appear more compelling to them based on their more limited technical knowledge. In this case, whilst the detailed evidence gathered by the advanced civilization may be compelling, their efforts to communicate are wasted. 

In order to maximise the chance of success the more advanced civilization must play by a set of rules that are comprehensible to the less advanced civilization. If the goal is to communicate with any kind of SETI-capable civilization, rather than one that is technologically similar, SETI civilizations should select targets on the basis that the targets could be aware of basic outward evidence of their own existence. This leads to the third law of mutual detectability:

\begin{itemize}
\item {\em 3. Opportunity} -- The subset of symmetrical evidence used to select a target that is based on intrinsic signal properties forms the {\em common denominator information} (CDI). Key to incentivizing both parties is the use of evidence that both can recognize. Communication chances are therefore enhanced when parties choose similar or matching types of CDI. 
\end{itemize}

An optimal strategy is one where substantial information on habitability may be gleaned by both parties from conceptually simple data. I argue that one good example of this is the observation of transiting HZ exoplanets located within the ETZ. The transit method is conceptually very simple yet opens up the possibility of determining whether the location, internal composition and atmosphere of a planet is conducive to habitation. Indeed transit observations may even show evidence of bio- or even techno-signatures. The ETZ is currently being surveyed by the Breakthrough Listen project \citep{2020AJ....160...29S} and I return to this specific situation in more detail in Section~\ref{transit}.

There is of course no way to ensure that both civilizations choose the same or similar form of CDI. For our own situation on Earth, information may be available beyond our knowledge that might indicate our existence to another civilization. Unless their planet has simpler information that is accessible to us and indicates their existence, we would be less motivated to target them, reducing the chances of communication. The Law of Opportunity argues that the odds are improved when both civilizations are employing comparable CDI which, if kept simple, casts the SETI ``net'' as wide as possible. There is a game-theory advantage even for advanced civilizations to base their targeting strategy on the most basic evidence, if their goal is to detect any SETI-capable civilization. The use of simple evidence increases the likelihood for the  mutual recognition of mutual evidence, which is key to incentivizing both parties. 

The remaining obstacle to establishing communication is the question of which party should transmit and which should observe. Of course, the optimal solution is that both parties should do both. However, from our own experience, it is easy to believe that both parties may pursue the ``cautious'' approach of observing, relying on the other to transmit. The SETI Paradox \citep{2006physics..11283Z} highlights the failure of this approach as a universal strategy. Under the principle of mutual detectability both parties can assess which of them has the upper hand in terms of having access to better quality CDI. This motivates the fourth and final law of mutual detectability:

\begin{itemize}
    \item {\em 4. Superiority} -- The onus to transmit falls on the party with superior CDI about the other party.
\end{itemize}

Adherence to the principle of mutual detectability breaks the SETI Paradox due to the unequal incentive among the two parties. The party with access to better CDI will have the greater incentive to initiate contact, assuming an initially equal desire to succeed. Of course, they may still refrain out of a fear of the unknown consequences of success. But the fact the incentive is not shared equally means that, other factors being equal (an implicit premise of the SETI Paradox), they are more likely to overcome their reticence. The fourth law therefore places the ``onus to transmit'' upon them.

It is important to stress the very limited and specific meaning of ``superiority'' here. It is based only upon the quality of CDI. It therefore does not include other information that the more advanced of the two civilisations may also have access to. It should also depend only on the strength of the intrinsic information signal and not on the balance of different detector technologies that may be available to either party. To see how these principles can shape a specific strategy we now consider the case of transiting systems.

\section{Application to transiting planets in the Earth's transit zone} \label{transit}

By way of example of how to apply the principle of mutual detectability in Section~\ref{sec:mutual} we consider here the question of whether or not we should engage in METI with transiting HZ planets that are located within the ETZ \citep[c.f.][]{2016AsBio..16..259H}. This provides us and any potential observer on the target planet with the ability to obtain mutual evidence of our mutual existence, in accordance with the Law of Mutuality. It also provides both parties with symmetric information, in accordance with the Law of Symmetry.

The available information content of transits is governed by the time-integrated number of host photons blocked by the planet, which I refer to as the transit {\em signal strength}. For a planet of radius $R_{\rm p}$ with period $P$ the observed signal strength integrated for a fixed time interval covering multiple transits scales as
   \be
   S \propto F_* (R_{\rm p}/R_*)^{2} (t_{14}/P), ~~ F_* \propto L_*/D^2, \label{strength}
   \ee
where $F_*$, $L_*$ and $R_*$ are the host star flux, luminosity and radius, $t_{14}$ is the transit duration and $D$ is the distance from the observer.

A civilization on a transiting planet located within the ETZ can similarly use Eq.~(\ref{strength}) to determine the signal strength of Earth's transit. Noting that $D$ is common we can express the relative {\em intrinsic} signal strength as:
  \be
  \frac{S}{S_{\oplus}} = \frac{L_*}{L_{\odot}} \left( \frac{R_*}{R_{\odot}}\right)^{-2} \left( \frac{R_{\rm p}}{R_{\oplus}}\right)^2 \left( \frac{P}{\mbox{yr}} \right)^{-1} \frac{t_{14}}{t_{\oplus}}, \label{rel-strength}
  \ee
where $t_{\oplus} = 12.9$~hours is the Earth's transit duration as viewed by a distant observer located on the Ecliptic plane. Using the standard convention of expressing the host light blocked by the transiting planet in units of parts-per-million (ppm), the denominator of Eq.~(\ref{rel-strength}) equates to an intrinsic time-integrated transit signal strength for Earth of $S_{\oplus} = 10^3~L_{\odot}$~ppm~hours~yr$^{-1}$. 

Eq~\ref{rel-strength} provides a fundamental measure of the  intrinsic strength of a transit signal compared to $S_{\oplus}$. This quantity satisfies the definition of CDI stated in the Law of Opportunity in Section~\ref{sec:mutual} in that it is independent of the observer's technology and encodes information on the habitability of a planet using basic data that should be understood by any civilisation able to conduct transit surveys. As mentioned in Section~\ref{sec:mutual}, there is no guarantee that another civilization would choose CDI along the lines of Eq~(\ref{rel-strength}); there are many other possible measures of habitability. Instead of using the light blocked by the planet they may consider instead the potential light blocked by the presence of an atmosphere surrounding a planet, which would scale as $R_{\rm p}$ rather than $R_{\rm p}^2$ \citep[e.g.][]{Morgan2019}. This would constitute similar CDI so would likely not compromise the chances of communication. But, they may instead employ a different measure based on the chemistry of a planet's atmosphere, or something else entirely. The Law of Opportunity states that to incentivize both parties the form of CDI should be mutually recognizable. The simpler the evidence the greater the chance of its recognition by both parties. In the case of transiting planets Eq~(\ref{rel-strength}) is about as simple as it gets and therefore SETI-capable civilizations wishing to cast their search net as wide as possible would, no matter how advanced their own capabilities, be motivated to adopt CDI of a form that is not too dissimilar.

On the basis of Eq~(\ref{rel-strength}), transiting systems viewed from Earth with intrinsic transit strengths $S < S_{\oplus} = 10^3~L_{\odot}$~ppm~hours~yr$^{-1}$ provide less information for us than we provide for them. The Law of Superiority then implies that they have the onus to transmit. This is a decision that both they and we can evaluate and agree upon. In the event that both parties use very different forms of CDI it may be that there is, by chance, still agreement as to who has the onus to transmit. But there may be disagreement, resulting in both or neither party transmitting. The use of comparable CDI therefore reduces the chance that no transmission occurs.

To illustrate the consequences of using  Eq.(\ref{rel-strength}) as CDI, I consider main sequence stars that are assumed to obey scaling relations between luminosity, mass and radius of the form
\be
\frac{L_*}{L_{\odot}} = a \left( \frac{M_*}{M_{\odot}} \right)^{\alpha}, ~~ 
\frac{R_*}{R_{\odot}} = b \left( \frac{M_*}{M_{\odot}} \right)^{\beta},
\label{scales}
\ee
where $a, \alpha, b$ and $\beta$ may take different values for different ranges of host mass $M_*$. These relations imply
   \be
   \frac{M_*}{M_{\odot}} = c \left( \frac{L_*}{L_{\odot}} \right)^{\gamma}, ~~~ \frac{R_*}{R_{\odot}} = d \left( \frac{L_*}{L_{\odot}} \right)^{\delta}, \label{lum-rels}
   \ee
where $c = a^{-1/\alpha}$, $\gamma =  1/\alpha$, $d = b \alpha^{-\beta/\alpha}$ and $\delta = \beta/\alpha$. Specific values for these constants are given in Table~\ref{host-props} for A0, G2 (Sun-like) and M0 stars, based on the piece-wise linear relations in \cite{2018MNRAS.479.5491E}. The values in Table~\ref{host-props} allow for reasonable interpolation to other hosts of neighbouring class.

\begin{deluxetable*}{ccccccccccccc}
\tablenum{1}
\tablecaption{Relations of stellar mass $M_*$, radius $R_*$ , luminosity $L_*$ and main-sequence lifetime $\tau_*$ used in Section~\ref{transit} for stars of A0, G2 and M0 type, based on the piece-wise fits presented by \cite{2018MNRAS.479.5491E}. The coefficients $a,b,c,d$ and power-law indices $\alpha, \beta, \gamma, \delta$ can be used in Eq~(\ref{scales}) and (\ref{lum-rels}) to interpolate to neighbouring spectral types.}
\label{host-props}
\tablewidth{0pt}
\tablehead{
\colhead{Type} & \colhead{$M_*/M_{\odot}$} &
\colhead{$L_*/L_{\odot}$} & \colhead{$R_*/R_{\odot}$} & \colhead{$\tau_*/$Gyr} & \colhead{$a$} & \colhead{$\alpha$} &
\colhead{$b$} & \colhead{$\beta$} &
\colhead{$c$} & \colhead{$\gamma$} & 
\colhead{$d$} & \colhead{$\delta$}
}
\startdata
A0 & 2.1 & 25 & 2.2 & 0.084 & 1.023 & 4.329 & 1.410 & 0.609 & 0.995 & 0.231 & 1.406 & 0.141 \\
G2 & 1.0 & 1.0 & 1.0 & 10 & 0.984 & 5.743 & 1.000 & 1.355 & 1.003 & 0.174 & 1.004 & 0.236 \\
M0 & 0.6 & 0.08 & 0.6 & 76 & 0.791 & 4.572 & 1.000 & 1.005 & 1.053 & 0.219 & 1.053 & 0.220 \\
\enddata
\end{deluxetable*}

\begin{figure}[ht!]
\plotone{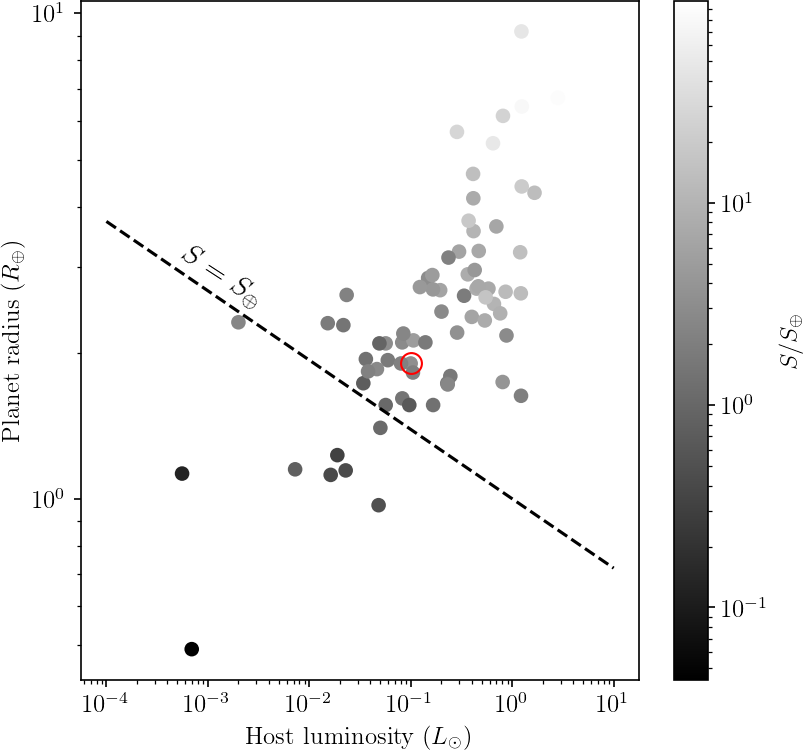}
\caption{The relative intrinsic transit signal strength, as defined by Eq~(\ref{rel-strength}), for 74 exoplanets with $T_{\rm eq}$ within 30\% of that of Earth ($T_{\oplus} = 255$~K). Only one of these planets, K2-155~d (circled) lies within the ETZ. The dashed line is the threshold of Eq~(\ref{onus}) that determines the onus to transmit. Above this line we have better CDI and therefore more incentive (onus) to transmit. Civilizations on planets below the line would have game-theory incentive to transmit to us. There are currently no confirmed transiting planets below the line with $T_{\rm eq}$ within 30\% of $T_{\oplus}$ and that lie within the ETZ, though there are probably a thousand or more such systems to be found. Data is obtained from the NASA Exoplanet Archive. \label{fig:sig}}
\end{figure}

For planets in the HZ both $P$ and $t_{14}$ can be related directly to $L_*$. Assuming a Bond albedo not very different from Solar System planets, the planet equilibrium temperature $T_{\rm eq}$ can be expressed as
  \be
  \frac{T_{\rm eq}}{T_{\oplus}} \simeq \left( \frac{L_*}{L_{\odot}} \right) ^{1/4} \left( \frac{a_{\rm p}}{\mbox{au}} \right) ^{-1/2}, \label{teq}
  \ee
where $T_{\oplus} = 255$~K is the Earth equilibrium temperature and $a_{\rm p}$ is the planet semi-major axis. From Kepler's third law we know that $P^2 \propto a_{\rm p}^3 M_*^{-1}$, which, combined with Eq~(\ref{teq}) and the mass-luminosity relation in Eq~(\ref{lum-rels}), leads to
   \be
   \frac{P}{\mbox{yr}} \approx c \left( \frac{L_*}{L_{\odot}} \right)^{\gamma+3/4} \label{period}
   \ee
for planets that lie within the HZ (i.e. having $T_{\rm eq} \simeq T_{\oplus}$). Using Eq~(\ref{teq}) with $T_{\rm eq} = T_{\oplus}$ and the relations of Eq~(\ref{lum-rels}) and (\ref{period}), the transit duration is characteristically
   \be
   t_{14} \simeq \frac{P R_*}{\pi a_{\rm p}} \approx cd \, t_{\oplus} \left( \frac{L_*}{L_{\odot}} \right)^{\gamma+\delta+1/4}. \label{duration}
   \ee
Substituting Eqns~(\ref{duration}) and (\ref{period}) into Eq~(\ref{rel-strength}), together with the radius-luminosity relation in Eq~(\ref{lum-rels}), leads to
   \be
   \frac{S}{S_{\oplus}} \approx \frac{1}{d} \left( \frac{L_*}{L_{\odot}} \right) ^{(1-2\delta)/2} \left( \frac{R_{\rm p}}{R_{\oplus}} \right) ^2. \label{rs-lum}
   \ee
When $S < S_{\oplus}$ observers located on HZ planets within the ETZ have access to superior CDI on us than we do on them. According to the Law of Superiority, this game-theory advantage means that they have the onus to transmit. Eq~(\ref{rs-lum}) shows that this is the case when
   \be
   \frac{R_{\rm p}}{R_{\oplus}} \lesssim d^{1/2} \left( \frac{L_*}{L_{\odot}} \right) ^{(2\delta-1)/4} \approx \left( \frac{L_*}{L_{\odot}} \right)^{-1/7}. \label{onus}
   \ee
The condition of Eq~(\ref{onus}) shows that the value of $R_{\rm p}$ at which both parties have equal CDI depends entirely on the stellar radius-luminosity relation in Eq~(\ref{lum-rels}). The range of $d$ and $\delta$ in Table~\ref{host-props} is sufficiently small that the right-hand expression in Eq~(\ref{onus}) is valid across all hosts from M through to A-type stars. Table~\ref{host-props} also shows that $R_*$ has a shallow dependence on $L_*$, leading to a remarkably shallow relation between $R_{\rm p}$ and $L_*$ in Eq~(\ref{onus}). However, host luminosity varies by a factor of 300 from M0 to A0 hosts, so there is still a factor two variation in $R_{\rm p}$  for hosts within this range.

A potential concern with using a CDI parameter like $S$, based on detectability, is that it may be of most relevance only for a brief time in the evolution of a civilisation's technology. For example, a little over two decades ago we did not know of any exoplanetary system. At the current time we know of more than 4,000 exoplanets and in only a decade or so we can expect this haul to increase by perhaps another order of magnitude with surveys in progress or in planning. Given this explosive growth in discovery, it is very easy to imagine that a civilisation that is ahead of us technologically by just a century or so may know about essentially all exoplanetary systems in its neighbourhood. They may choose to pursue game theory approaches based on factors other than detectability. For example, they may focus instead on the age of hosts as a proxy for the level of advancement of potential target civilisations. 

A counter-argument to this concern is that it is difficult to conceive that civilisations would continue targeted SETI programs over timescales vastly in excess of the timescale for discovering exoplanets; these timescales are surely strongly coupled. A targeted-SETI strategy motivated by game theory considerations has to focus on civilisations that are most likely to be still engaged in SETI. Whilst there is no way for us to determine what subset of technological civilisations are SETI-engaged, CDI that is framed on exoplanet detectability provides a sensible basis for a targeted-SETI strategy optimised for civilisations who are most likely to be SETI-active.

The above considerations could prompt us to consider whether there is game theory advantage to focus on planets around younger stars, perhaps with ages similar to our Sun. Perhaps younger stars are more likely to host relatively young technological civilisations like us that are SETI-engaged. In reality we simply have no information on which to base such a judgement. There is no obvious reason why the typical longevity of a technological civilisation or, perhaps of greater relevance, the typical timescale over which they engage in SETI, should correlate with the age of their host star. Older hosts may be as likely to host a succession of relatively short-lived technological civilisations as they are to host a few longer-lived ones. In either case the typical timescale over which civilisations engage in SETI may not correlate with their longevity. There is therefore no obvious candidate for a game-theory motivated targeted-SETI strategy based on host age.

However, it is important for us to consider how the host lifetime affects the likelihood that a technological civilisation has the chance to become established at all. Using the mass-luminosity relation in Eq~(\ref{lum-rels}), the stellar main-sequence lifetime can be estimated as
   \be
   \tau_* = 10~\mbox{Gyr}~\left( \frac{M_*}{M_{\odot}} \right)\left(\frac{L_*}{L_{\odot}}\right)^{-1} \simeq 10\, c~\mbox{Gyr}~\left( \frac{L_*}{L_{\odot}} \right)^{\gamma-1} \approx 10~\mbox{Gyr}~\left( \frac{L_*}{L_{\odot}} \right)^{-4/5}.
   \label{lifetime}
   \ee
If we consider that $\sim 1$~Gyr is perhaps the minimum time required for a SETI-capable civilization to appear then Eq~(\ref{lifetime}) indicates that we should not expect SETI civilisations for hosts with $L_* \gtrsim 20~L_{\odot}$, where the limit corresponds to an A1 host. At this limit Eq~(\ref{onus}) indicates that the onus to transmit falls on us for any HZ planet larger than $0.8~R_{\oplus}$. For solar luminosity hosts the onus is clearly evenly split at Earth-sized planets, whilst for a civilization on an HZ planet orbiting a M0 host ($L_* = 0.08~L_{\odot}$) the onus to transmit falls on them when $R_{\rm p} \lesssim 1.5~R_{\oplus}$. This limit happens to correspond to the lower bound of the Fulton gap \citep{2017AJ....154..109F}.

Whilst we have game-theory incentive to transmit to super-solar luminosity F and A-type stars hosting transiting Earth-analogues in the ETZ, such hosts are comparatively rare. Assuming a Kroupa form for the stellar initial mass function \citep{2001MNRAS.322..231K}, super-solar mass stars are around 10 times rarer than sub-solar mass stars above the hydrogen-burning limit. Furthermore, exoplanet occurrence studies show that HZ Earth-analogues around M-type stars may be significantly more common than around more luminous hosts. \cite{Bryson_2020} find an occurrence rate of $0.015^{+0.011}_{-0.007}$ planets per host for planets around GK-type stars with orbital period and radius within 20\% of Earth's. Using a different selection pipeline, \cite{2015ApJ...807...45D} find an occurrence rate of $0.16^{+0.17}_{-0.07}$  for HZ planets of radius 1--1.5~$R_{\oplus}$ around M-type stars. Whilst significant arguments have been put forward against sustainable life being hosted around M dwarf stars, recent research indicates that it may be possible for planets within the HZ of M-dwarfs stars to host long-lived oceans of water \citep{2020MNRAS.496.3786M}. Some aspects of M-dwarf environments may even be advantageous for supporting life \citep{2016PhR...663....1S}.  

Taking current evidence of the planet occurrence rate in combination with the stellar mass function, it appears that HZ Earth-like planets are far more likely to be found around sub-solar hosts. Any SETI-capable life on such systems have more game-theoretic motivation to engage in METI with us than vice-versa, making the SETI Paradox an unstable equilibrium solution from a game-theory perspective, even if we choose to remain silent.

Figure~\ref{fig:sig} shows the relative intrinsic transit signal strength, $S/S_{\oplus}$, for 74 confirmed exoplanets that have $T_{\rm eq}$ within 30\% of $T_{\oplus} = 255$~K. Only one of these systems, K2-155~d \citep{2018AJ....155..124H,2018MNRAS.476L..50D}, is located within the ETZ. K2-155~d is a potentially habitable super-Earth orbiting a bright M-dwarf host that lies just $0.^{\circ}16$ south of the Ecliptic plane at a distance of around 60~pc from Earth. Using the measured stellar and planet parameters from Tables~1 and 2 of \cite{2018MNRAS.476L..50D}, the relative signal strength for K2-155~d is $S/S_{\oplus} = 3.3$, which means that we have better CDI about it than a civilization on K2-155~d would have about Earth. In this case the game-theory onus falls on us to transmit to K2-155~d. However, the absence of HZ planets with $S < S_{\oplus}$ within the ETZ is almost entirely due to the fact that this region has not yet been thoroughly surveyed for transiting planets around sub-solar luminosity hosts. Around 60\% of the ETZ will be surveyed for transiting planets as part of the extended NASA TESS mission.

\cite{2016AsBio..16..259H} used a Galactic model to calculate that we should expect up to around $10^5$ potential K and G-type hosts to be located within the ETZ, and even more than this if M-dwarf stars are also included. Using Eq~(\ref{scales}) and (\ref{teq}), for Earth-analogue planets (i.e. having $R_{\rm p} \simeq R_{\oplus}$ and $T_{\rm eq} \simeq T_{\oplus}$), the probability that their orbit is orientated to yield a transit signal from Earth is
   \be
   \mbox{P}_{\rm T}\mbox{(HZ)} = \frac{R_{\rm p} + R_*}{a_{\rm p}} \simeq \frac{R_*}{a_{\rm p}} = \frac{R_{\odot}}{\mbox{au}} \left( \frac{R_*}{R_{\odot}} \right) \left( \frac{a_{\rm p}}{\mbox{au}} \right)^{-1} \simeq 4.6 \times 10^{-3} \, d \left( \frac{L_*}{L_{\odot}} \right)^{\delta-1/2}.  \label{prob}
   \ee
So, even with $\mbox{P}_{\rm T}\mbox{(HZ)} = {\cal O}(10^{-2})$, we can expect transiting HZ planets within the ETZ to number potentially into the thousands. 

\cite{2016MNRAS.459.1233K} have argued that transits also offer a natural Schelling point in terms of the timing of a signal transmitted by a SETI-capable civilisation. They point out that civilisations could, for example, use laser emission timed to coincide with the onset of the transit signal itself, knowing that the transit epoch is likely to be when another civilisation might be observing them. From a game-theory perspective, an ETZ transit survey for HZ Earth analogues around low-luminosity hosts should be considered a high priority for targeted SETI searches. 

\section{Discussion} \label{discuss}

I have presented the principle of mutual detectability as a basis for enhancing the chance of success for targeted SETI surveys based on game theory motivation. Mutual detectability also highlights that primary evidence of habitability is not expected to be of equal quality to all parties. Some will be more incentivized than others, knowing that they have superior evidence, to send messages (engage in METI). 

The mutual detectability principle is embodied by four laws of mutuality, symmetry, opportunity and superiority. The first three laws highlight the enhanced probability of establishing contact when two SETI-capable parties become aware of mutual evidence of their mutual existence. By careful selection of evidence they can quantify which of them likely has access to better common evidence, so-called common denominator information (CDI). The game theory incentive rests with the civilization that has superior CDI and the fourth law states that the onus is therefore on them to transmit. The SETI Paradox, raising the spectre of all civilizations remaining collectively silent, seems an unlikely outcome given the imbalance of game-theory incentive. 

I have illustrated how transiting habitable-zone (HZ) planets located within the Earth Transit Zone (ETZ) form a subset for targeted SETI programs that is consistent with mutual detectability. I show that SETI-capable civilizations on Earth-like planets around sub-solar luminosity stars (planets that are known to be common) have a game theory incentive to initiate contact, and therefore an ``onus'' to transmit. K2-155~d is identified as a confirmed transiting HZ super-Earth that lies within the ETZ. However, game theory considerations suggest that the onus to transmit falls on us for this particular system. 

It is already recognised by the exoplanet community that HZ planets around low-luminosity hosts are common. They are typically good targets for transiting exoplanet searches as the host stars are small and so HZ Earth-sized planets have relatively large transit depth and are on relatively short period orbits, with the TRAPPIST-1 system being a nearby example \citep{2017Natur.542..456G}. Given the potential value of such systems to SETI searches I argue the case for undertaking a deep transit search over the ETZ with a view to assembling a catalogue of HZ Earth-analogue transits around sub-solar luminosity hosts. This sample would establish a catalogue of systems suitable for targeted SETI monitoring. Any SETI-capable civilization on these systems would have game-theory incentive to transmit to us. The direction of the line of nodes between the Ecliptic and Galactic planes is potentially a good starting point for such a survey, due to the relatively high concentration of stars.

Mutual transits are the most obvious example of a sample that satisfies mutual detectability. But it will certainly not be unique. For example, future direct imaging surveys of astrometrically-detected Earth-like planets would provide substantial habitability information. The mutual detectability principal can be used to define CDI for how these systems would likewise detect us directly, together with our astrometric signature. Whilst transit samples within the ETZ are not strongly favoured for us to transmit to, it may be that other samples outside the ETZ that are selected for mutually detectability may put a somewhat stronger onus on us to transmit. 

Success in SETI may rest upon us and other civilisations recognising that it may be best to make different choices for who to listen to and who to message.

\acknowledgements

This work is funded by a STFC-NARIT Newton Fund award. EK is grateful to the anonymous referee for useful comments.


\bibliography{SETI}{}

\begin{thebibliography}{}
\expandafter\ifx\csname natexlab\endcsname\relax\def\natexlab#1{#1}\fi
\providecommand{\url}[1]{\href{#1}{#1}}
\providecommand{\dodoi}[1]{doi:~\href{http://doi.org/#1}{\nolinkurl{#1}}}
\providecommand{\doeprint}[1]{\href{http://ascl.net/#1}{\nolinkurl{http://ascl.net/#1}}}
\providecommand{\doarXiv}[1]{\href{https://arxiv.org/abs/#1}{\nolinkurl{https://arxiv.org/abs/#1}}}

\bibitem[{Bryson {et~al.}(2020)Bryson, Coughlin, Batalha, Berger, Huber, Burke,
  Dotson, \& Mullally}]{Bryson_2020}
Bryson, S., Coughlin, J., Batalha, N.~M., {et~al.} 2020, Astron. J., 159, 279,
  \dodoi{10.3847/1538-3881/ab8a30}

\bibitem[{Cortellesi(2020)}]{2020arXiv200601167C}
Cortellesi, T. 2020, arXiv e-prints, arXiv:2006.01167.
\newblock \doarXiv{2006.01167}

\bibitem[{Deeg \& Belmonte(2017)}]{Wright2017}
Deeg, H.~J., \& Belmonte, J.~A., eds. 2017, {Exoplanets and SETI} (Cham:
  Springer International Publishing), 1--9,
  \dodoi{10.1007/978-3-319-30648-3_186-1}

\bibitem[{{D\'{i}ez Alonso} {et~al.}(2018){D\'{i}ez Alonso}, {Su{\'{a}}rez
  G{\'{o}}mez}, {Gonz{\'{a}}lez Hern{\'{a}}ndez}, {Su{\'{a}}rez
  Mascare{\~{n}}o}, {Gonz{\'{a}}lez Guti{\'{e}}rrez}, Velasco,
  Toledo-Padr{\'{o}}n, {de Cos Juez}, \& Rebolo}]{2018MNRAS.476L..50D}
{D\'{i}ez Alonso}, E., {Su{\'{a}}rez G{\'{o}}mez}, S., {Gonz{\'{a}}lez
  Hern{\'{a}}ndez}, J., {et~al.} 2018, MNRAS, 476, L50,
  \dodoi{10.1093/mnrasl/sly040}

\bibitem[{Dogaru(2019)}]{Dogaru2019}
Dogaru, A. 2019, {MSc dissertation, Univ. Manchester}.
\newblock \url{https://bit.ly/33kPZFv}

\bibitem[{Dressing \& Charbonneau(2015)}]{2015ApJ...807...45D}
Dressing, C.~D., \& Charbonneau, D. 2015, Astrophys. J., 807, 45,
  \dodoi{10.1088/0004-637X/807/1/45}

\bibitem[{Eker {et~al.}(2018)Eker, Bak\i\textcommabelow~s, Bilir, Soydugan,
  Steer, Soydugan, Bak\i\textcommabelow~s, Ali{\c{c}}avu\textcommabelow~s,
  Aslan, \& Alpsoy}]{2018MNRAS.479.5491E}
Eker, Z., Bak\i\textcommabelow~s, V., Bilir, S., {et~al.} 2018, MNRAS, 479,
  5491, \dodoi{10.1093/mnras/sty1834}

\bibitem[{Filippova {et~al.}(1991)Filippova, Kardashev, Likhachev, \&
  Strelnitskj}]{10.1007/3-540-54752-5_225}
Filippova, L.~N., Kardashev, N.~S., Likhachev, S.~F., \& Strelnitskj, V.~S.
  1991, in Bioastronomy, ed. J.~Heidmann \& M.~J. Klein (Berlin, Heidelberg:
  Springer Berlin Heidelberg), 254--258

\bibitem[{Fulton {et~al.}(2017)Fulton, Petigura, Howard, Isaacson, Marcy,
  Cargile, Hebb, Weiss, Johnson, Morton, Sinukoff, Crossfield, \&
  Hirsch}]{2017AJ....154..109F}
Fulton, B.~J., Petigura, E.~A., Howard, A.~W., {et~al.} 2017, Astron. J., 154,
  109, \dodoi{10.3847/1538-3881/aa80eb}

\bibitem[{Gertz(2016)}]{2016JBIS...69...31G}
Gertz, J. 2016, J. Br. Interplanet. Soc., 69, 31.
\newblock \doarXiv{1605.05663}

\bibitem[{Gillon {et~al.}(2017)Gillon, Triaud, Demory, Jehin, Agol, Deck,
  Lederer, de~Wit, Burdanov, Ingalls, Bolmont, Leconte, Raymond, Selsis,
  Turbet, Barkaoui, Burgasser, Burleigh, Carey, Chaushev, Copperwheat, Delrez,
  Fernand~es, Holdsworth, Kotze, {Van Grootel}, Almleaky, Benkhaldoun, Magain,
  \& Queloz}]{2017Natur.542..456G}
Gillon, M., Triaud, A.~H., Demory, B.-O., {et~al.} 2017, \nat, 542, 456,
  \dodoi{10.1038/nature21360}

\bibitem[{Heller \& Pudritz(2016)}]{2016AsBio..16..259H}
Heller, R., \& Pudritz, R.~E. 2016, Astrobiology, 16, 259,
  \dodoi{10.1089/ast.2015.1358}

\bibitem[{Hirano {et~al.}(2018)Hirano, Dai, Livingston, Fujii, Cochran, Endl,
  Gand~olfi, Redfield, Winn, Guenther, Prieto-Arranz, Albrecht, Barragan,
  Cabrera, Cauley, Csizmadia, Deeg, Eigm{\"{u}}ller, Erikson, Fridlund, Fukui,
  Grziwa, Hatzes, Korth, Narita, Nespral, Niraula, Nowak, P{\"{a}}tzold, Palle,
  Persson, Rauer, Ribas, Smith, \& {Van Eylen}}]{2018AJ....155..124H}
Hirano, T., Dai, F., Livingston, J.~H., {et~al.} 2018, Astron. J., 155, 124,
  \dodoi{10.3847/1538-3881/aaaa6e}

\bibitem[{Kipping \& Teachey(2016)}]{2016MNRAS.459.1233K}
Kipping, D.~M., \& Teachey, A. 2016, MNRAS, 459, 1233,
  \dodoi{10.1093/mnras/stw672}

\bibitem[{Kroupa(2001)}]{2001MNRAS.322..231K}
Kroupa, P. 2001, MNRAS, 322, 231, \dodoi{10.1046/j.1365-8711.2001.04022.x}

\bibitem[{Lacki {et~al.}(2020)Lacki, Brzycki, Croft, Czech, DeBoer, DeMarines,
  Gajjar, Isaacson, Lebofsky, MacMahon, Price, Sheikh, Siemion, Drew, \&
  Worden}]{2020arXiv200611304L}
Lacki, B.~C., Brzycki, B., Croft, S., {et~al.} 2020, arXiv e-prints,
  arXiv:2006.11304.
\newblock \doarXiv{2006.11304}

\bibitem[{Lin {et~al.}(2014)Lin, {Gonzalez Abad}, \&
  Loeb}]{2014ApJ...792L...7L}
Lin, H.~W., {Gonzalez Abad}, G., \& Loeb, A. 2014, Astrophys. J. Lett., 792,
  L7, \dodoi{10.1088/2041-8205/792/1/L7}

\bibitem[{Moore \& Cowan(2020)}]{2020MNRAS.496.3786M}
Moore, K., \& Cowan, N.~B. 2020, MNRAS, 496, 3786,
  \dodoi{10.1093/mnras/staa1796}

\bibitem[{Morgan {et~al.}(2019)Morgan, Kerins, Awiphan, McDonald, Hayes,
  Komonjinda, Mkritchian, \& Sanguansak}]{Morgan2019}
Morgan, J., Kerins, E., Awiphan, S., {et~al.} 2019, MNRAS, 486,
  \dodoi{10.1093/mnras/stz783}

\bibitem[{Pinchuk {et~al.}(2019)Pinchuk, Margot, Greenberg, Ayalde, Bloxham,
  Boddu, {Gerardo Chinchilla-Garcia}, Cliffe, Gallagher, Hart, Hesford,
  Mizrahi, Pike, Rodger, Sayki, Schneck, Tan, Xiao, \&
  Lynch}]{2019AJ....157..122P}
Pinchuk, P., Margot, J.-L., Greenberg, A.~H., {et~al.} 2019, Astron. J., 157,
  122, \dodoi{10.3847/1538-3881/ab0105}

\bibitem[{Price {et~al.}(2020)Price, Enriquez, Brzycki, Croft, Czech, DeBoer,
  DeMarines, Foster, Gajjar, Gizani, Hellbourg, Isaacson, Lacki, Lebofsky,
  MacMahon, de~Pater, Siemion, Werthimer, Green, Kaczmarek, Maddalena, Mader,
  Drew, \& Worden}]{2020AJ....159...86P}
Price, D.~C., Enriquez, J.~E., Brzycki, B., {et~al.} 2020, Astron. J., 159, 86,
  \dodoi{10.3847/1538-3881/ab65f1}

\bibitem[{Sheikh {et~al.}(2020)Sheikh, Siemion, Enriquez, Price, Isaacson,
  Lebofsky, Gajjar, \& Kalas}]{2020AJ....160...29S}
Sheikh, S.~Z., Siemion, A., Enriquez, J.~E., {et~al.} 2020, Astron. J., 160,
  29, \dodoi{10.3847/1538-3881/ab9361}

\bibitem[{Shields {et~al.}(2016)Shields, Ballard, \&
  Johnson}]{2016PhR...663....1S}
Shields, A.~L., Ballard, S., \& Johnson, J.~A. 2016, \physrep, 663, 1,
  \dodoi{10.1016/j.physrep.2016.10.003}

\bibitem[{Shostak(2004)}]{2004IAUS..213..409S}
Shostak, S. 2004, in Bioastronomy 2002 Life Among Stars, ed. R.~Norris \&
  F.~Stootman, Vol. 213, 409

\bibitem[{Siemion {et~al.}(2013)Siemion, Demorest, Korpela, Maddalena,
  Werthimer, Cobb, Howard, Langston, Lebofsky, Marcy, \&
  Tarter}]{2013ApJ...767...94S}
Siemion, A.~P., Demorest, P., Korpela, E., {et~al.} 2013, Astrophys. J., 767,
  94, \dodoi{10.1088/0004-637X/767/1/94}

\bibitem[{Wlodarczyk-Sroka {et~al.}(2020)Wlodarczyk-Sroka, Garrett, \&
  Siemion}]{2020MNRAS.tmp.2570W}
Wlodarczyk-Sroka, B.~S., Garrett, M., \& Siemion, A. 2020, MNRAS,
  \dodoi{10.1093/mnras/staa2672}

\bibitem[{Wright(2020)}]{2020arXiv200801817W}
Wright, J.~T. 2020, arXiv e-prints, arXiv:2008.01817.
\newblock \doarXiv{2008.01817}

\bibitem[{Zaitsev(2006)}]{2006physics..11283Z}
Zaitsev, A. 2006, arXiv e-prints, physics/0611283.
\newblock \doarXiv{physics/0611283}

\bibitem[{Zhang {et~al.}(2020)Zhang, Werthimer, Zhang, Cobb, Korpela, Anderson,
  Gajjar, Lee, Li, Pei, Zhang, Huang, Wang, Zhu, Duan, Zhang, Jin, Zhu, \&
  Li}]{2020ApJ...891..174Z}
Zhang, Z.-S., Werthimer, D., Zhang, T.-J., {et~al.} 2020, Astrophys. J., 891,
  174, \dodoi{10.3847/1538-4357/ab7376}

\end{thebibliography}



\end{document}